\documentclass[aps,twocolumn,graphicx,superscriptaddress,preprintnumbers,nofootinbib]{revtex4-1}
\usepackage{amsmath,amssymb,bbm}
\usepackage{graphicx}
\usepackage{color,hyperref}
\hypersetup{colorlinks,breaklinks,
            linkcolor=blue,urlcolor=blue,
            anchorcolor=blue,citecolor=blue}
\usepackage{multirow}

\def\roughly#1{\mathrel{\raise.3ex\hbox
{$#1$\kern-.75em\lower1ex\hbox{$\sim$}}}}

\newcommand{\pslash}{D\kern-0.15em\raise0.17ex\llap{/}\kern0.15em\relax}

\newcommand{\gerda}      {{\sc Gerda}}
\newcommand{\phase}      {{\sc Phase}}
\newcommand{\majorana}   {{\sc Majorana}}
\newcommand{\majoranadem}{{\sc Majorana Demonstrator}}
\newcommand{\LSGe}       {{\sc LSG}e}

\newcommand{\nubb}   {\ensuremath{0\nu\beta\beta}}
\newcommand{\nubbhl} {\ensuremath{T_{1/2}^{0\nu}}}
\newcommand{\mee} {\ensuremath{|m_{ee}|}}
\newcommand{\Qbb}    {\ensuremath{Q_{\beta\beta}}}
\newcommand{\ctsper} {\ensuremath{\text{cts}/(\text{keV}\cdotp\text{kg}\cdotp\text{yr})}}
\newcommand{\kgyr} {\ensuremath{\text{kg}\cdotp\text{yr}}}
\newcommand{\isot}[2]{\ensuremath{^{\text{#2}}}\text{#1}}

\newcommand{\sep}{ & }

\begin{document}
\title{Probing flavor models with \isot{Ge}{76}-based experiments on neutrinoless double-$\beta$ decay}
\author{Matteo Agostini}
\email{matteo.agostini@ph.tum.de}
\affiliation{Physik Department and Excellence Cluster Universe, Technische Universit\"at M\"unchen, Germany}
\affiliation{Gran Sasso Science Institute (INFN), L'Aquila, Italy}
\author{Alexander Merle}
\email{amerle@mpp.mpg.de}
\affiliation{Max-Planck-Institut f\"ur Physik (Werner-Heisenberg-Institut),
%F\"ohringer Ring 6, 80805 M\"unchen, 
Germany}
\author{Kai Zuber}
\email{zuber@physik.tu-dresden.de}
\affiliation{Institute for Nuclear and Particle Physics, Technische Universit\"at Dresden, Germany}
\date{\today}

\begin{abstract}
The physics impact of a staged approach for double-$\beta$ decay experiments based
on \isot{Ge}{76} is studied.
The scenario considered relies on realistic time schedules envisioned by the
\gerda\ and the \majorana\ collaborations, which are jointly working towards the
realization of a future larger scale \isot{Ge}{76} experiment.
Intermediate stages of the experiments are conceived to perform 
quasi background-free measurements, and different data sets can be 
reliably combined to maximize the physics outcome.
The sensitivity for such a global analysis is presented, with focus on how 
neutrino flavor models can be probed already with preliminary phases of the
experiments.
The synergy between theory and experiment yields strong benefits for both sides:
the model predictions can be used to sensibly plan the experimental stages, and
results from intermediate stages can be used to constrain whole groups of
theoretical scenarios.
This strategy clearly generates added value to the experimental efforts, while
at the same time it allows to achieve valuable physics results as early as possible.
\end{abstract}

\preprint{MPP--2015--111}
%\pacs{}
%\keywords{}
\maketitle

%%%%%%%%%%%%%%%%%%%%%%%%%%%%
%\paragraph{Introduction}
%%%%%%%%%%%%%%%%%%%%%%%%%%%%

Neutrino physics led to big discoveries in the past decades, the greatest being
the observation of neutrino oscillations~\cite{Nu-osc}, which prove that
neutrino masses (albeit tiny) must be non-zero and that neutrino flavors mix.
In a nutshell, this means that an electron neutrino does not have a fixed mass but
it is rather a quantum-mechanical superposition of several mass eigenstates.
While nowadays most oscillation parameters are
known~\cite{Gonzalez-Garcia:2014bfa} and a new era of precision neutrino physics
has started, several fundamental questions are still unanswered. Probably the
most important question is whether neutrinos have a Majorana nature, i.e., if
they are identical to their antiparticles, which would signal a violation of
lepton number and thus lead beyond the very successful standard model of particle
physics. 
Such questions can be answered by the observation of neutrinoless double-$\beta$
decay (\nubb)~\cite{Furry:1939qr}, a nuclear transition in which two neutrons
decay simultaneously into two protons by emitting two electrons but no neutrino, 
thus changing lepton number by two units
and possibly signaling a Majorana neutrino mass~\cite{SV}.

\begin{figure}
 \includegraphics[width=1\columnwidth]{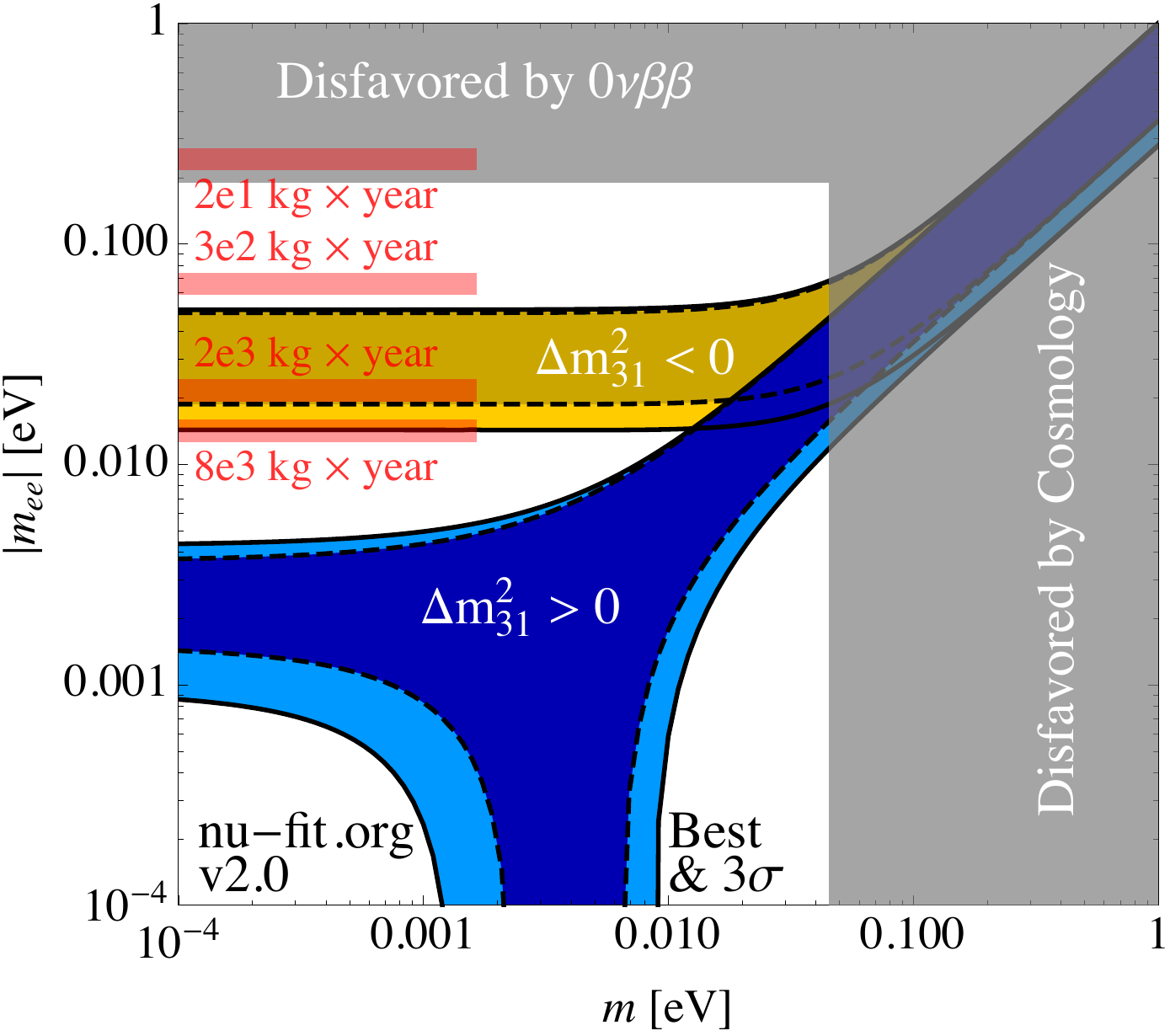}
 \caption{\label{fig:mee}%
    Most general regions for the effective mass \mee, calculated using global fit
    parameters (v2.0) from {\tt nu-fit.org}~\cite{Gonzalez-Garcia:2014bfa}, along
    with four specific values of sensitivity considered in the text. The broadening of the bands is due to
   nuclear physics uncertainties. The
    disfavored regions are the most optimistic bounds from \isot{Xe}{136}-based
    experiments~\cite{Albert:2014awa,Gando:2012zm}
    and Planck~\cite{Ade:2015xua}, the latter converted to the smallest neutrino
    mass and averaged between both mass orderings.}
\end{figure}

The experimental search for \nubb\ is a very active field of particle and
nuclear physics.
Various isotopes for which \nubb\ is energetically allowed and many detection techniques are
pursued. Examples are: \isot{Ge}{76}  with high purity Ge
detectors%~\cite{Ackermann:2012xja,Abgrall:2013rze}
, \isot{Te}{130} with TeO$_2$
bolometric detectors~\cite{Alfonso:2015wka}, \isot{Xe}{136} with liquid Xe time
projection chambers~\cite{Albert:2014awa}, or Xe-loaded organic liquid
scintillator detectors~\cite{Gando:2012zm}. Historically, \isot{Ge}{76}-based
experiments have been leading the field, and the resulting constraints on the
half-life of the process are among the most stringent
ones~\cite{Gunther:1997ai,Aalseth:2002rf,Agostini:2013mzu}.
Two \isot{Ge}{76}-based experiments are currently active and will yield results in
the near future: \gerda~\cite{Ackermann:2012xja} and
\majorana~\cite{Abgrall:2013rze}. These two collaborations conceive of
eventually realizing a common large scale \isot{Ge}{76} (\LSGe)
experiment~\cite{NLDBD},
capable of probing the
theoretically allowed region for the inverted mass ordering (i.e., the
experimentally favorable scenario corresponding to the upper yellow band in
\figurename~\ref{fig:mee}).
For such a challenging experiment, a highly modular
design and a staged approach implementation are needed, 
meaning that the target
mass will be progressively increased.

This paper presents realistic projections of the sensitivity achievable by a
global analysis of the data from current and future experiments searching for
\nubb\ in \isot{Ge}{76}. 
Among \nubb\ experiments, the ones based on \isot{Ge}{76} stand out because they
are designed to perform quasi background-free measurements.
Their data can hence be combined without limiting assumptions on the background
modeling. 
We also point out that the sensitivity of a global analysis should be considered when
planning the mass-increasing strategy of a project, in order to maximize the
benefit for both theory and experiment.  
Indeed, in case no signal will be observed, large classes of theoretical
neutrino models can be excluded already by intermediate stages of an
experiment.

%%%%%%%%%%%%%%%%%%%%%%%%%%%%
%\paragraph{\nubb\ and neutrino mass sum rules}
%%%%%%%%%%%%%%%%%%%%%%%%%%%%

The physics observable accessible with \nubb\ experiments is the
\emph{effective Majorana neutrino mass} $\mee = |m_1 c_{12}^2 c_{13}^2 + m_2
s_{12}^2 c_{13}^2 e^{i\alpha_{21}} + m_3 s_{13}^2 e^{i(\alpha_{31} -
2\delta)}|$, which depends on sines ($s$) and cosines ($c$) of the leptonic mixing
angles $\theta_{ij}$, the mass eigenvalues, and the phases~\cite{Lindner:2005kr}.
It is related to the $0\nu\beta\beta$ half-life by~\cite{Smolnikov:2010zz}:
\begin{equation}
 1/ \nubbhl  = G_{0\nu} |\mathcal{M}_{0\nu}|^2 \mee^2,
 \label{eq:mee}
\end{equation}
where $G_{0\nu} = 2.42\cdotp 10^{-26}\ {\rm yr}^{-1}{\rm eV}^{-2}$ is a phase-space factor and
$\mathcal{M}_{0\nu}$ is the dimensionless
nuclear matrix element (NME) which parametrizes the nuclear physics involved.
The allowed range for \mee\ as a function of the smallest neutrino mass $m$
is constrained by the experimental measurements of the neutrino mixing
parameters, see \figurename~\ref{fig:mee}. 
Nevertheless, information about the absolute neutrino mass that is inferred by
combining all experimental information are affected by systematic uncertainties
of the analysis procedure~\cite{Maneschg:2008sf}, the NMEs~\cite{NMEs},
and the mixing
parameters~\cite{Lindner:2005kr}.
Consequently, even pinning down the neutrino mass ordering -- whether
normal, $m_1 < m_2 < m_3$ (blue), or inverted, $m_3 < m_1 < m_2$ (yellow) -- is
challenging. 

\begin{figure}
   \includegraphics[width=1.\columnwidth]{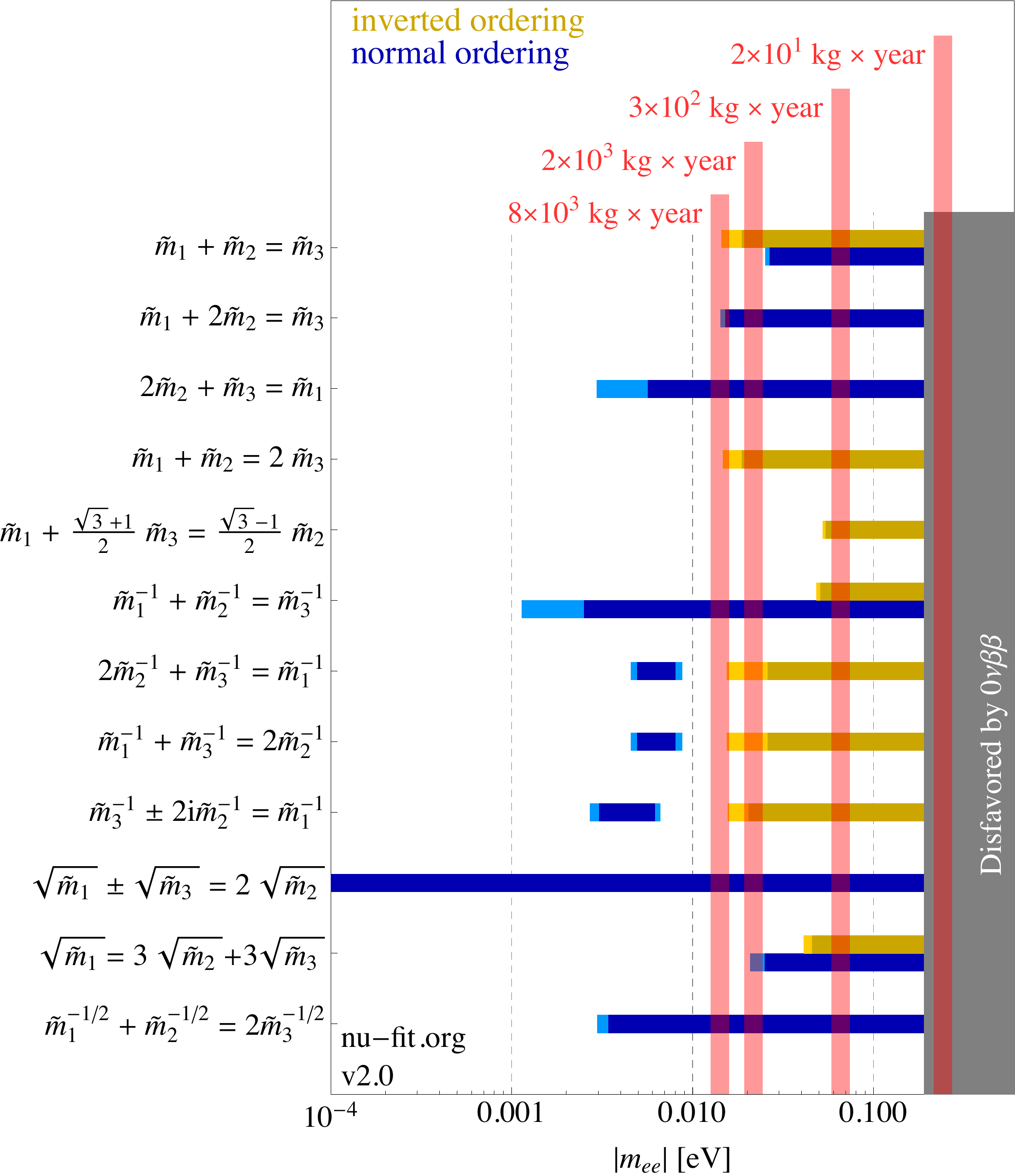}
   \caption{\label{Fig:Bands}%
   Range of the $0\nu\beta\beta$ effective mass allowed for different classes of
   neutrino flavor models that are characterized by respective sum rules.  The
   sensitivity for different stages of \isot{Ge}{76}-based experiments
   considered in the text is also displayed.}
\end{figure}

This situation could drastically change with additional input from neutrino physics.
The smallness of neutrino masses can be theoretically explained 
by suppression mechanisms at tree-~\cite{tree-seesaws} or loop-level~\cite{loop-seesaws}
and the large mixing angles by flavor models based on discrete
symmetries~\cite{Flavour-Reviews}, which explain them by relating their values
to properties of finite symmetry groups. 
While many of those models yield
similar predictions for the accessible observables 
-- so that their experimental distinction is prevented unless the precision is increased by
about two orders of magnitude --
certain classes of models predict clear
\emph{correlations} between observables. Prime examples are \emph{neutrino mass
sum rules}~\cite{Sum-Rules,Our-Sum-Rules}, such as $\tilde m_1 + \tilde m_2 =
\tilde m_3$ or $1/\tilde m_1 + 1/\tilde m_3 = 2/\tilde m_2$, which correlate the
complex neutrino mass eigenvalues $\tilde m_i$. These rules are complex
equations and thus deliver \emph{two} pieces of information: a constraint on the
mass scale $m$ and some relation between the Majorana phases $\alpha_{21,31}$.
The most extensive study available~\cite{Our-Sum-Rules} investigated more than
50~flavor models divided into 12~classes, which -- as
\figurename~\ref{Fig:Bands} shows  -- can
greatly decrease the allowed range for \mee, thereby offering the possibilities
of gaining valuable knowledge on the neutrino sector already by the intermediate
steps in a staged approach towards detecting \nubb.
%

%%%%%%%%%%%%%%%%%%%%%%%%%%%%
%\paragraph{\isot{Ge}{76}-based experiments}
%%%%%%%%%%%%%%%%%%%%%%%%%%%%

The advantages of using high purity Ge (HPGe) detectors for the \nubb\ search have
been recognized early~\cite{Fiorini:1967in}. 
HPGe detectors can be produced from germanium isotopically enriched in
\isot{Ge}{76} (\isot{Ge}{enr}, typically 87\% enrichment). % from 7.8\% natural abundance). 
The experimental signature expected for \nubb\ inside the detector is a peak in
the energy spectrum at the $Q$-value of the \isot{Ge}{76} decay
($\Qbb=2039.061(7)$\,keV~\cite{Mount:2010zz}). Remarkable advantages of this detection
technique are the intrinsic radio-purity of the detectors, the excellent
spectroscopic performance ($\lesssim$0.1\% energy resolution at \Qbb), and the
high detection efficiency. 
In addition, these detectors are a well consolidated technology 
widely used for $\gamma$-ray spectroscopy, which proved to be reliable and suitable
for long-term experiments.  
The detector geometries considered for \nubb\ experiments include three types:
\emph{coaxial}, \emph{Broad Energy Germanium} (BEGe), and \emph{P-type Point Contact} (PPC)~\cite{Agostini:2013loa,Mertens:2015hwa}. 
Each geometry results in a specific electric field inside the
detector, which affects the performance of event-reconstruction techniques based on
the time evolution of the read-out electrical signals (i.e., pulse shape analysis).
HPGe detectors must be operated at cryogenic temperatures and are commonly
installed in vacuum cryostats.
This approach was adopted also by past \nubb\
experiments~\cite{Aalseth:2002rf,Gunther:1997ai}, which operated coaxial-type
detectors in ultra-low background cryostats surrounded by massive lead and
copper shieldings.

Nowadays,
the \majorana\ collaboration is pursuing a design based on PPC-type detectors
and multiple cryostat modules built from ultrapure electroformed copper.
Two modules are currently being assembled
(i.e., the \majoranadem~\cite{Abgrall:2013rze})
at the Sanford Underground Research Facility in Lead, South Dakota
(USA). The first module hosts 16.8\,kg of \isot{Ge}{enr} detectors and will be
fully operational in the second half of 2015. The completion of the second
cryostat containing further 12.6\,kg of \isot{Ge}{enr} detectors is scheduled by
the end of 2015.
The experiment is designed to operate the detectors at a background level of
$0.75\cdotp10^{-3}$\,\ctsper\ at \Qbb.\footnote{
   The design goal of the \majoranadem\ is typically quoted as
   3\,cts/$(\text{ton}\cdotp\text{yr})$ in a region of interest of 4\,keV.  
}
The \gerda\ collaboration is instead exploring
an alternative design in which an array of bare \isot{Ge}{enr}
detectors is operated directly in ultra radio-pure liquid argon, which acts as
coolant material, passive shielding against the external radioactivity, and
active veto-system when its scintillation light is detected. The setup is
installed in the Gran Sasso underground laboratories of INFN in
Italy. \gerda\ has recently completed its first phase of
operation (\phase~I), during which $\sim$15 kg of \isot{Ge}{enr} detectors
(mostly of coaxial type) 
have been operated with a background level of $10^{-2}$\,\ctsper,
yielding a limit of $\smash{\nubbhl}\geq 2.1\cdotp10^{25}$\,yr
\cite{Agostini:2013mzu}. The apparatus is currently being upgraded to operate
additional 17\,kg of \isot{Ge}{enr} BEGe-type detectors
and new sensors for the argon scintillation light. A second data taking phase
(\phase~II) is planed to start in the second half of 2015 with a background level of
$10^{-3}$\,\ctsper\, at \Qbb~\cite{Majorovits:2015gla}.

The \majoranadem\ and \gerda\ \phase~II will together start the exploration of
\nubbhl\ at the scale of $10^{26}$\,yr, i.e., $\mee \sim 0.1$\,eV. 
The results collected by the two experiments during the first years of operation
will be essential to define the design of the \LSGe\ experiment and down-select
the best technologies to operate $\gtrsim1000$\,kg of target mass at a
background level of $\lesssim10^{-4}$\,\ctsper\ at \Qbb.
With such parameters, the \LSGe\ experiment will probe \nubbhl\ sensitivity at
the level of $10^{27}$--$10^{28}$\,yr and hence explore
an essential part of the parameter space
allowed for inverted mass ordering or -- with a fortunate value of $\theta_{12}$
and better precision on that parameter coming from experiments like
JUNO~\cite{He:2014zwa} or
RENO-50~\cite{Kim:2014rfa} -- even the whole parameter space.

%%%%%%%%%%%%%%%%%%%%%%%%%%%%
%\paragraph{Global sensitivity}
%%%%%%%%%%%%%%%%%%%%%%%%%%%%
The
sensitivity achievable by a global analysis of \gerda~\phase~I and \phase~II,
the \majoranadem, and a future \LSGe\ experiment has been studied by
assuming the data sets listed in \tablename~\ref{tab:datasets}. 
Following the
analysis approach adopted by the \gerda\ collaboration, data from \gerda~\phase~I are
divided into two data sets according to the two types of detectors operated.\footnote{
   The data set correspond to the ``golden'' and ``BEGe'' data sets of
   Ref.~\cite{Agostini:2013mzu}. A third data set considered in the analysis of the
   collaboration (the ``silver'' data sets, about 6\% of the overall exposure) is
   not considered here, due to its negligible contribution to the overall
   sensitivity of the experiment.
}
The separation into two data sets is assumed also
for \phase~II. The experimental parameters such as efficiencies, background
level and duration are taken from the published
values~\cite{Agostini:2013mzu,Agostini:2015pta}. The energy resolution is taken
from the most recent R\&D
results~\cite{Agostini:2015pta,Agostini:2014hra,Majorovits:2015gla}.
BEGe-type detectors will provide higher energy resolution and
superior background reduction performance with respect to the coaxial type. Data
from \majoranadem\ are also split between the two modules into two data sets.
Efficiencies of PPC- and BEGe-type detectors are assumed to be equal. This assumption
is fully consistent with the first results presented by the
\majorana\ collaboration~\cite{Mertens:2015hwa}, and from which the energy resolution is
taken.
\begin{table}
   \caption{\label{tab:datasets}%
   Parameters assumed for each data set: detector mass, efficiency $\epsilon$,
   background level at \Qbb, energy resolution (full width at half maximum,
   FWHM) at \Qbb, start time, and duration of the data taking $\Delta t$.
   The start time of the current (future) experiments is indicated with
   $t_0$ ($t_1$) and expected to be in the second half of 2015 (in the 2020s).
}
   \begin{ruledtabular}
   \begin{tabular}{lcc c c cc@{}}
      ~data
      & ~  mass ~
      & ~ $\epsilon$ ~
      & background
      & FWHM
      & start 
      & $\Delta t$\\

      ~~set
      & [kg]
      & 
      & \multirow{2}{*}{$\left[\dfrac{\text{cts}}{\text{keV}\cdotp\text{kg}\cdotp\text{yr}}\right]$}
      & [keV]
      & time
      & [yr] \\

      \\

      \hline
      \multicolumn{5}{l}{\gerda~\phase~I:}  \\
      \emph{coaxial} & 12.2 & 0.62 & $1.1\cdotp10^{-2}$ & 4.4 & Nov\,2011\sep1.3 \\
      \emph{BEGe}    & 2.8  & 0.66 & $0.5\cdotp10^{-2}$ & 2.9 & ~Jul\,2012\sep0.8 \\
      \hline                   
      \multicolumn{5}{l}{\gerda~\phase~II:}  \\
      \emph{coaxial} & 17.7 & 0.62 & $  1\cdotp10^{-3}$   & 4.0 & $t_0$\sep4\\
      \emph{BEGe}    & 20.0 & 0.65   & $  1\cdotp10^{-3}$   & 2.5 & $t_0$\sep4\\
      \hline                                      
      \multicolumn{5}{l}{\majoranadem:}  \\
      \emph{mod1} & 16.8 & 0.65 & $0.8\cdotp10^{-3}$ & 3.0 & $t_0$\sep4 \\
      \emph{mod2} & 12.6 & 0.65 & $0.8\cdotp10^{-3}$ & 3.0 & $t_0$+0.5\,yr\sep4 \\
      \hline                                      
      \multicolumn{5}{l}{Future large scale (\LSGe) experiment:}  \\
      \emph{mod1} & 200  & 0.65 & $  1\cdotp10^{-4}$   & 2.5 & $t_1$\sep10 \\
      \emph{mod2} & 200  & 0.65 & $  1\cdotp10^{-4}$   & 2.5 & $t_1$+1\,yr\sep9 \\
      \emph{mod3} & 200  & 0.65 & $  1\cdotp10^{-4}$   & 2.5 & $t_1$+2\,yr\sep8 \\
      \emph{mod4} & 200  & 0.65 & $  1\cdotp10^{-4}$   & 2.5 & $t_1$+3\,yr\sep7 \\
      \emph{mod5} & 200  & 0.65 & $  1\cdotp10^{-4}$   & 2.5 & $t_1$+4\,yr\sep6 \\
   \end{tabular}
   \end{ruledtabular}
\end{table}
A staged approach is assumed for the \LSGe\ experiments. Given
realistic constraints on the production of \isot{Ge}{enr} material\footnote{
   The Svetlana Department facility can currently deliver
   80-100\,kg of \isot{Ge}{76} per year~\cite{Agostini:2014hra}. 
   We realistically assume that the production can be doubled in the next years and
   with some investments coming from the \LSGe\ experiment.
}
the total target mass of 1000\,kg is
assumed to be progressively increased by installing one new module with  200\,kg
of detectors per year. The detectors are considered to perform similarly to
BEGe-type detectors. 
%The beginning of the data taking for this experiment is
%set to 2021. 
%Note that the parameters of each data set -- in particular start and stop time
%-- are realistic but partially arbitrary. 

The total number of \nubb\ events in each data set as a function of \nubbhl\ is
given by:
\begin{equation}
 N^{0\nu} = \ln 2 \cdot N_{A} \cdot \epsilon \cdot \eta/(m_{a} \cdot \nubbhl),
 \label{eq:conversion}
\end{equation}
where $N_A$ is the Avogadro's number, $\epsilon$ the efficiency, $\eta$ the
exposure, and $m_a$ the molar mass of \isot{Ge}{enr}. In this work, the exposure
$\eta$ is defined as the product of total detector mass and data taking time.
The efficiency $\epsilon$ is given by the product of four contributions: the
fraction of \isot{Ge}{76} in the detectors material ($\sim$87\%), the fraction
of the detector volume which is active (87\% for coaxial, 92\% for BEGe/PPC
detectors), the efficiency of the analysis cuts (90\%, dominated by pulse shape
analysis cuts), and the probability that \nubb\ events in the detector active
volume are correctly reconstructed at energy \Qbb\ (92\% for coaxial, 90\% for
BEGe/PPC detectors). These efficiencies are taken from
Ref.~\cite{Agostini:2013mzu}.  A duty cycle of 95\% is assumed for all
experiments, which accounts for the time needed to calibrate the detectors and
for ordinary hardware maintenance.

A statistical approach is adopted to estimate the \nubbhl\ lower limit
achievable by a global analysis of the various data sets. More than $10^{6}$
time-stamps are randomly selected. Given a time-stamp,
background events with a uniform energy distribution in the range
$\Qbb\pm0.1$\,MeV are generated with Monte
Carlo techniques. Events are generated independently  for each data set
according to its background level, exposure, and efficiency. A simultaneous fit
of all data sets is hence performed, using a constant probability density
function for the background and a Gaussian function for the \nubb\ signal (with
centroid at \Qbb). The 90\%~C.L.\ upper limit on number of \nubb\ counts
extracted from the fit is converted into a 90\%~C.L.\ lower limit on \nubbhl\ by
using Eq.~\eqref{eq:conversion}.

The fit procedure is based on an unbinned profile likelihood analysis in which
the number of \nubb\ counts is bounded to positive values. The free parameters
of the fit are the number of signal counts (the parameter of interest) and the
background levels (nuisance parameters). Systematic uncertainties (energy scale,
resolution, and efficiency) have been studied by adding Gaussian pull terms in
the likelihood function and found to worsen the limits by $\lesssim 1\%$. To
reduce the computational time, these systematic uncertainties have not been
included in the final simulation, as their effect is small for a limit-setting
experiment. The coverage of the method has been tested for a sample of
time-stamps and found to provide a conservative overcoverage. 

\begin{figure}
   \includegraphics[angle=270,width=\columnwidth]{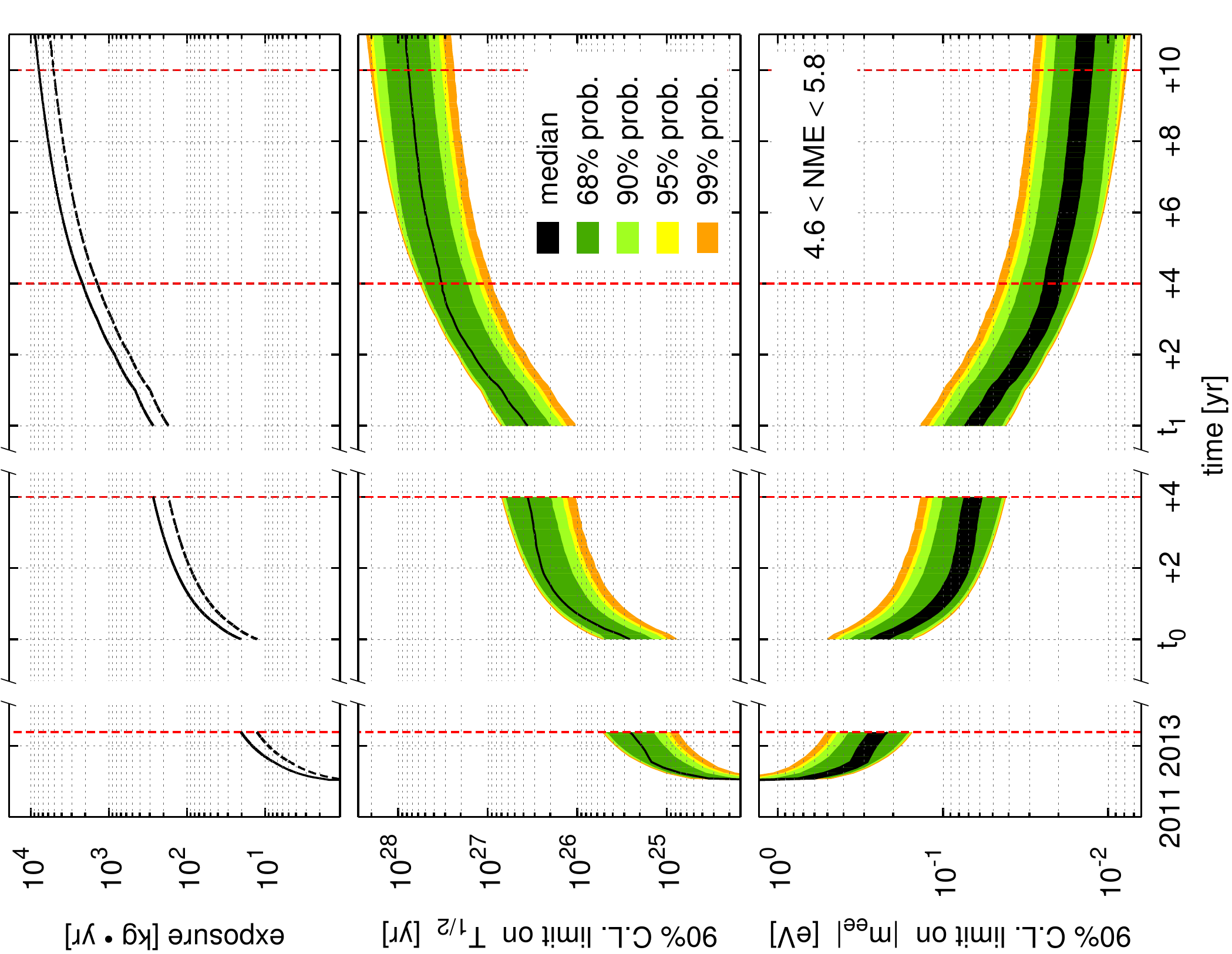}
   \caption{\label{fig:sensitivity}%
   Top panel: integrated exposure assumed for
   the calculation as a function of time before (solid line) and after (dashed
   line) efficiency correction. Middle (bottom) panel: distribution of the
   90\%~C.L.\ lower limits on \nubbhl\  (upper limits on \mee) derived by a
   global analysis of multiple realizations of the experiments. The
   vertical red lines corresponds to specific values of exposure/sensitivity
   discussed in the text. 
   The time axis is broken and future dates are given with respect to the start
   of \gerda~\phase~II and the \majoranadem\ ($t_0$), and of the future \LSGe\
   experiment ($t_1$).}
\end{figure}

The results of these computations are shown in \figurename~\ref{fig:sensitivity}. 
The top panel illustrates the integrated
exposure over time.
%The exposure is shown for the detector mass and for the actual \isot{Ge}{76}
%target mass after \emph{all} efficiency corrections.
%Three segments of exposure increase are shown, corresponding to periods in which at
%least one experiment is running.
The increase of exposure is driven by 
\gerda~\phase~I (between 2012 and mid 2013), \gerda~\phase~II and the
\majoranadem\ (between $t_0$ and $t_0$+4\,yr) and the \LSGe\ experiment (between
$t_1$ and $t_1$+10\,yr).
The middle panel shows the distribution of the 90\%~C.L.\ lower limits on
\nubbhl. 
%The median of the distribution is shown along with the central
%intervals for different probabilities. 
The uppermost part of the distribution is
populated by the data set realizations with no background events at \Qbb\ (i.e.,
fully background-free). It grows linearly with the exposure and has a sharp
cut-off due to the constraint $N^{0\nu}\geq 0$ imposed on the fit. The bottom
panel shows the distribution of the 90\% upper limits on the effective mass
\mee. This distribution is computed by converting each \nubbhl\ limit through
Eq.~\eqref{eq:mee}, which introduces an additional systematic uncertainty on
each limit due to the uncertain NME calculations. The effect of this systematic
uncertainty is maximally included in the plot assuming NME values in the range
between 4.6 and 5.8~\cite{Our-Sum-Rules}.\footnote{%
   The range covers all the calculations available in the literature
   except for the shell model, which predicts an outlier value for NME of 2.3~\cite{Our-Sum-Rules}.
   Expanding the range up to the shell model prediction increases our upper
   bounds on \mee\ by a factor of 2.}
Thus, the median line becomes a band and the central intervals broaden.

\figurename~\ref{fig:sensitivity} shows how successive experiments can be designed
to improve the median experimental sensitivity by an order of magnitude. 
\gerda~\phase~I reached a sensitivity of 
   $\nubbhl > 2.6\cdotp10^{25}$\,yr ($\mee > 215$--$272$\,meV)
with an exposure of $\sim$20\,\kgyr\ (corresponding to the
leftmost red line in \figurename~\ref{fig:sensitivity}). \gerda~\phase~II and
the \majoranadem\ will improve the sensitivity up to 
%
   %$\nubbhl > 3.7\cdotp10^{26}$\,yr ($\mee > 58$--$73$\,meV)
   $\nubbhl > 4\cdotp10^{26}$\,yr ($\mee > 58$--$74$\,meV)
by collecting an exposure of
%
   %$3.1\cdotp10^{2}$\,\kgyr\ 
   $3\cdotp10^{2}$\,\kgyr\ 
in 4\,yr. The \LSGe\ experiment will
finally rise the sensitivity up to 
%
   %$\nubbhl > 7.8\cdotp10^{27}$\,yr ($\mee > 13$--$16$\,meV) 
   $\nubbhl > 8\cdotp10^{27}$\,yr ($\mee > 13$--$16$\,meV) 
in 10\,yr of data taking and a final exposure of
%
   %$7.9\cdotp10^{3}$\,\kgyr. 
   $8\cdotp10^{3}$\,\kgyr. 
It is noteworthy that a sensitivity of  
%
   %$\nubbhl > 3.3\cdotp10^{27}$\,yr ($\mee > 19$--$25$\,meV)
   $\nubbhl > 3\cdotp10^{27}$\,yr ($\mee > 19$--$24$\,meV)
can be reached with about 
%
   %$2.2\cdotp10^{3}$\,~\kgyr\ 
   $2\cdotp10^{3}$\,~\kgyr\ 
in 4 years of data taking with the \LSGe\ experiment.

The ultimate question to answer is what can be learned about neutrino physics from
future \isot{Ge}{76}-based experiments.
\figurename~\ref{fig:mee} 
shows how challenging will be to fully probe the parameter space allowed
for \mee\ in the most general situation, even considering the future \LSGe\
and inverted mass ordering. 
Intermediate sensitivity stages seem not to be able to provide
remarkable physics results unless a positive signal is observed. 
However, we demonstrate in \figurename~\ref{Fig:Bands} that whole groups of
more specific neutrino flavor models, namely those which predict particular mass
sum rules, can be excluded already by intermediate stages.
For example,
the sum rule $\tilde m_1^{-1} + \tilde m_2^{-1} = \tilde m_3^{-1}$ yields for
inverted ordering a smallest allowed neutrino mass of 51\,meV (48\,meV)
for the best-fit (3$\sigma$) values of the neutrino mass squares.
This region can be almost probed by \gerda~\phase~II and the \majoranadem, and
fully probed (i.e., even with \emph{all} uncertainties) by first stages of the \LSGe\
experiment.
Thus, by using the sum rule predictions as
orientation when planning the stages, one can exploit the
synergies between model predictions and experimental sensitivities to greatly
enhance the physics outcome even of the intermediate stages. 
This synergy goes so far that some groups of models could be distinguished in
spite of the uncertainties involved, and our considerations would be
strengthened further by a better knowledge on the NMEs, on the neutrino mass ordering,
or on the mixing angle $\theta_{12}$ -- and even more by the observation that
the sum rule predictions are quite stable in what regards certain types of
theoretical (radiative) corrections~\cite{SR-RGE}. 
Additionally, we would like to point out the a remarkable number of models could
be already ruled out with $\sim2\cdotp10^{3}$\,\kgyr\ of exposure.

Such an exposure could be collected 
%with 200\,kg of detector mass and 10\,yr of data taking, i.e., 
by a single module of the \LSGe\
experiment or by upgrades of \gerda~\phase~II and the 
\majoranadem\ which are already under consideration within the experimental
community~\cite{LNGS2020, NLDBD}.

In conclusion, realistic sensitivity projections have been presented for the
current and future \isot{Ge}{76}-based experiments.
A global analysis of different data sets is reliable and should be performed.
The global sensitivity and its impact on flavor models should be
carefully considered when designing the mass-increasing strategy of the future
projects. 

Synergies between theory and experiment can push us to new
frontiers in neutrino physics, provided that we make proper use of them.
%
%%%%%%%%%%%%%%%%%%%%%%%%%%%%
\begin{acknowledgments}
%\emph{Acknowledgments} 
M.~A. would like to thank A.~Caldwell, J.~Detwiler, L.~Pandola, B.~Schwingenheuer,
and S.~Sch\"onert for valuable discussions.
A.~M. acknowledges partial support from the European Union FP7 ITN-INVISIBLES
(Marie Curie Actions, PITN-GA-2011-289442).
\end{acknowledgments}

\end{document}